\documentclass[final,5p,times,twocolumn]{elsarticle}
\usepackage{amsmath,latexsym,mathtools}
\usepackage{soul}
\usepackage[x11names,dvipsnames,table]{xcolor} 

\usepackage{multicol}
\usepackage{dcolumn}
\usepackage{subfig}

\usepackage[pscoord]{eso-pic}
\newcommand{\placetextbox}[3]{
\setbox0=\hbox{#3}
\AddToShipoutPictureFG*{
\put(\LenToUnit{#1\paperwidth},\LenToUnit{#2\paperheight}){\vtop{{\null}\makebox[0pt][c]{#3}}}%
}%
}%

\usepackage{makecell}
\usepackage{booktabs} 

\usepackage{etoolbox}
\usepackage{amsmath,latexsym,mathtools}
\usepackage{comment}
\usepackage{footnpag}
\usepackage{float} 
\usepackage{xcolor}    
\usepackage[multiple]{footnotehyper}
\usepackage{chngcntr}
\journal{Physics of the Dark Universe}
\counterwithout{footnote}{section} 

\newcommand{\correct}[1]{{\color{black} #1}}
\usepackage{slashed}
\usepackage{braket}
\usepackage{amssymb}
\usepackage{amsfonts}
\usepackage{float}
\usepackage[utf8]{inputenc}
\usepackage{amsmath}
\usepackage{physics}
\usepackage{graphicx} 
\usepackage{dcolumn} 
\usepackage{bm} 

\usepackage{subcaption}
\usepackage{changepage}
\usepackage{xcolor}
\usepackage{float} 
\journal{Physics of the Dark Universe}

\usepackage[colorlinks]{hyperref}
\usepackage[thinc]{esdiff}

\AtBeginDocument{\hypersetup{
	colorlinks=true,
	linkcolor=black!40!blue,
	citecolor=black!40!blue,
	filecolor=black,
	urlcolor=black!45!blue,
}}

\usepackage{comment}

\begin{document}

\title{Multi-field TDiff theories for cosmology}

\author[1]{Diego Tessainer \corref{cor1}}
\ead{dtessain@ucm.es}

\author[1]{Antonio L. Maroto}
\ead{maroto@ucm.es}

\author[1]{Prado Martín-Moruno}
\ead{pradomm@ucm.es}

\cortext[cor1]{Corresponding author}

\affiliation[1]{organization={Departamento de Física Teórica and Instituto de Física de Partículas y del Cosmos (IPARCOS-UCM)}, addressline={\\Universidad Complutense de Madrid},postcode={28040}, city={Madrid}, country={Spain.}}

\begin{abstract}
We consider theories which break the invariance under diffeomorphisms (Diff) down to transverse diffeomorphisms (TDiff) in the matter sector, consisting of multiple scalar fields. In particular, we regard shift-symmetric models with two free TDiff scalar fields in a flat Robertson-Walker spacetime (FLRW) and use the perfect fluid approach to study their dynamics. As a consequence of the symmetry breaking, an effective interaction between the fields is induced from the conservation of the total energy-momentum tensor, without the necessity to introduce any explicit interacting term in the Lagrangian. We study the different single-field domination regimes and analyze the energy exchange between the fields. Thereupon, we introduce an application of these models for the description of interactions in the dark sector, and compare the theoretical predictions of our model to observational data from Type Ia supernovae.
\end{abstract}


\maketitle

\placetextbox{0.85}{0.95}{IPARCOS-UCM-24-042}

\section{Introduction}
It is widely known, as observational data indicate, that our Universe currently exhibits an accelerated expansion \cite{Riess_1998}. Many models explain this as the consequence of a dark energy component dominating the cosmic expansion, taken to be a cosmological constant in the standard model. However, there are other alternatives, such as \textit{quintessence}, involving a canonical scalar field with a dynamical equation of state determined by its potential \cite{Tsujikawa:2013fta}, and \textit{k-essence} \cite{Armendariz-Picon:2000ulo}, which also display dynamical dark energy and can avoid fine-tuning problems, but include non-canonical kinetic terms in the action. Additionally, observational data also indicate that most of the matter composition of our Universe is dark matter \cite{Freese:2008cz}.  Furthermore, it is nowadays recognized that there exists a tension in the Hubble parameter $H_0$ measurements \cite{Riess:2019cxk,unknown,Abdalla:2022yfr}, which could be alleviated by models involving dark sector interactions \cite{DiValentino:2017iww} or phantom models \cite{DiValentino:2020naf}, in which the energy density of the dark energy component increases with the expansion. In addition, it has been proven that in order to ease both the $H_0$ and $S_8$ cosmological tensions simultaneously by taking into account new physics that is relevant only at late cosmic times, a dark energy component crossing the phantom limit is necessary \cite{Heisenberg:2022lob}.
On the other hand, as the nature of the dark sector is unknown, possible modifications of gravity at cosmological scales are often considered \cite{CANTATA:2021ktz}. Regarding this possibility, multiple modified gravity theories extending upon General Relativity (GR) have been explored \cite{Clifton:2011jh}.

Even if GR provides a very powerful tool for studying gravity and cosmology, theories breaking invariance under diffeomorphisms (Diff) have been recently gaining popularity, with one of the most prominent ones being Unimodular Gravity (UG) \cite{einsteinUG,Unruh.40.1048,Carballo-Rubio:2022ofy}. In UG, the metric determinant is taken to be a fixed non-dynamical field and the Diff invariance is broken down to transverse diffeomorphisms (TDiff) and Weyl rescalings. UG theories could provide a solution to the problem of vacuum-energy which does not gravitate in this type of theories \cite{articleEllis}. Nevertheless, in this work we will focus on theories that are only invariant under TDiff, which have lately started to be studied more deeply.
Thus for instance, TDiff models beyond UG have been studied in references \cite{Alvarez:2006uu,Pirogov:2011iq, Bello-Morales:2023btf,Bello-Morales:2024vqk}.  The cosmological evolution in TDiff-invariant theories propagating a scalar graviton mode was analyzed  in reference \cite{Bello-Morales:2023btf}. On the other hand, TDiff invariant models with broken diffeomorphisms in the matter sector   have been analyzed in references \cite{Maroto:2023toq,Jaramillo-Garrido:2023cor} for single scalar fields. There it is shown that even though on small scales such theories behave as standard Diff models, on super-Hubble scales the behaviour can be drastically different, thus opening up a 
wide range of possibilities for cosmological model building. Thus, in particular, a simple TDiff model for dark matter based on a free scalar field was proposed in references \cite{Maroto:2023toq,Jaramillo-Garrido:2023cor}. A unified TDiff model for the dark sector has been considered in reference \cite{Alonso-Lopez:2023hkx}. A general classification of 
single-field TDiff models based on their speed of sound  and equation of state was performed in reference \cite{Jaramillo-Garrido:2024tdv}. TDiff models for single abelian gauge fields can be found in reference \cite{Maroto:2024mkx} and their phenomenological implications for cosmic magnetic field 
evolution in reference \cite{Maroto:2024roe}.

In this work we will extend the previous works and consider multi-scalar TDiff invariant models in the matter sector in  flat Robertson-Walker (FLRW) spacetimes. We will specifically regard shift-symmetric models, which are invariant under shift transformations of the field, i.e., $\phi\rightarrow\phi+C$, where $C$ is a constant. Thus, we will only consider the exact kinetic domination regime for each field. The motivation behind this approach lies in the fact that in this way we can avoid fine-tuning problems depending on the specific choice for the potential term in the action. On the other hand, not considering any mass or potential-like terms in the action also results in the Einstein-Hilbert action only receiving higher-order radiative corrections, which also motivates our choice to only break the Diff symmetry in the matter sector.

Unlike the single-field case, the energy-momentum tensor (EMT) conservation will entail an effective interaction between the fields as a consequence of the symmetry breaking even without introducing any interaction terms in the Lagrangian. This fact opens up a wide range of phenomenological implications for multi-field models. Particularly, we will apply this effect to describe an interacting dark sector, comparing its predictions with observational data.

The work is organized as follows. In section \ref{secII} we briefly review the TDiff formalism, focusing on shift-symmetric theories and lay the groundwork for our particular models. Section \ref{secIII} is devoted to explain the theoretical framework for multi-scalar TDiff models. In section \ref{secIV} we perform a numerical analysis for our model, applying it to the dark sector. Results will be compared both with observations and $w$CDM, and physical predictions for our TDiff model will be obtained. Finally, in section \ref{secV} we will discuss the conclusions.

\section{Single-field shift-symmetric TDiff theories}\label{secII}
In this section we will briefly recap the main results obtained for shift-symmetric TDiff theories involving one scalar field. 
\subsection{Transverse diffeomorphisms and matter action}
Let us first consider a general infinitesimal coordinate transformation $x^\mu\mapsto x'^\mu=x^\mu+\xi^\mu(x)$ given by the vector field $\xi$. As it is well known, the variation  of metric tensor $g_{\mu\nu}(x)$ will be given by its Lie derivative, i.e.,
\begin{equation}
    \delta g_{\mu\nu}=\mathfrak{L}_\xi(g_{\mu\nu})=-\nabla_\nu\xi_\mu-\nabla_\mu\xi_\nu,
    \label{gmunu_lie}
\end{equation}
and thus it follows that the metric determinant ($g:=|\mathrm{det}(g_{\mu\nu})|$) will transform according to
\begin{equation}
    \delta g=gg^{\mu\nu}\delta g_{\mu\nu}=-2g\nabla_\mu\xi^\mu.
    \label{gdet_lie}
\end{equation}

Let us now write down our action. This is
\begin{equation}
    S=S_{\mathrm{EH}}[g_{\mu\nu}]+S_\mathrm{mat}[g_{\mu\nu},\phi],
    \label{gen_action_model}
\end{equation}
where $S_\mathrm{mat}$ denotes the matter part of the action involving a single scalar field $\phi$. Since we will only break the Diff symmetry in the matter action, the geometrical part will just be the usual Einstein-Hilbert action
\begin{equation}
    S_\mathrm{EH}[g_{\mu\nu}]=-\frac{1}{16\pi G}\int\dd^4x\sqrt{g}\,R.
    \label{SHE}
\end{equation}
On the other hand, the matter part will read
\begin{equation}
    S_\mathrm{mat}[g_{\mu\nu},\phi]=\int\dd^4x f(g)\,\mathcal{L}(g_{\mu\nu}(x),\phi(x),\partial_\mu\phi(x)),
\end{equation}
where $\mathcal{L}$ denotes the corresponding scalar under Diff Lagrangian density and $f(g)$ an arbitrary function of the metric determinant. Recalling \eqref{gmunu_lie} and \eqref{gdet_lie} we can compute $\delta_\xi S$, which, after integration by parts and assuming that the fields vanish at infinity, reads \cite{Maroto:2023toq}
\begin{equation}
    \delta_\xi S=\int\dd^4x\,\partial_\mu\xi^\mu\,[f(g)-2gf'(g)]\,\mathcal{L}.
    \label{deltaS}
\end{equation}
Thus, we see that the action is invariant under any infinitesimal coordinate transformation (Diff invariant) only when $f(g)-2gf'(g)=0$, i.e. $f(g)\propto\sqrt{g}$. However, the action is also invariant for any form of $f(g)$ if the transformations satisfy $\partial_\mu\xi^\mu=0$.
This corresponds to a smaller subgroup of symmetry, the transverse diffeomorphisms (TDiff).

\subsection{Single scalar-field models in the kinetic regime}
 Let us first consider the matter part of the action with a simple kinetic term \cite{Maroto:2023toq,Jaramillo-Garrido:2023cor}:
 \begin{equation}
    S_\mathrm{mat}=\int\dd^4x\,\frac{1}{2}f(g)\,\partial_\mu\phi\partial^\mu\phi,
     \label{Sm_1field}
 \end{equation}
 where $f(g)$ is a positive coupling function of the metric determinant. We consider this function to be positive-valued as the wrong sign for the kinetic term is typically related to a ghost instability \cite{Rubakov:2014jja}. The corresponding equation of motion reads
 \begin{equation}\label{eq: complete EoM for psi}
    \partial_\mu\big( f(g) \partial^\mu \phi \big) = 0 \,,
\end{equation}
 and the EMT will be defined as usual:
 \begin{equation}
    T^{\mu\nu}:=-\frac{2}{\sqrt{g}}\frac{\delta S_\mathrm{mat}}{\delta g_{\mu\nu}},
     \label{EMTgen}
 \end{equation}
 which in this case reads
 \begin{equation}
    T_{\mu\nu}=\frac{f(g)}{\sqrt{g}}\left(\partial_\mu\phi\partial_\nu\phi-F(g)\,g_{\mu\nu}\Box\phi\right),
     \label{EMT_1f_genspacetime}
 \end{equation}
 where we have defined $F(g):=\dd\ln f(g)/\dd\ln g$.
Since we are not modifying the Einstein-Hilbert action, the Bianchi identities are preserved and thus the local conservation of the EMT will still hold \cite{Maroto:2023toq,Jaramillo-Garrido:2023cor} under solutions of Einstein equations. 
 
 In relation to the background geometry, we will consider a spatially flat FLRW metric. Since we have less gauge freedom than in the Diff case, we will not generally be able to perform a coordinate change that fixes the lapse function to one and we will have more physical degrees of freedom than in the Diff case. Thus, our spacetime can be described by the following line element \cite{Alvarez:2007nn}:
 \begin{equation}
    ds^2=b^2(\tau)\,\dd\tau^2-a^2(\tau)\,\mathbf{d x}^2,
     \label{flat_RW}
 \end{equation}
 where $a(\tau)$ and $b(\tau)$ are the independent components that will act as the scale factor and lapse function, respectively; $\tau$ denotes the time-coordinate and $\mathbf{d x}$$^2$ corresponds to the spatial part of the spacetime metric. Both must be computed from Einstein equations\footnote{We will use the signature $(+,-,-,-)$ in this work.}. 

 Let us now apply the perfect fluid approach. It is worth recalling that, when $\partial_\mu\phi$ is a time-like vector, the EMT \eqref{EMT_1f_genspacetime} takes the form \cite{Jaramillo-Garrido:2023cor}
 \begin{equation}
    T_{\mu\nu}=(\rho+p)\,u_\mu u_\nu-p\,g_{\mu\nu},
     \label{pfluid_emt}
 \end{equation}
 where $\rho=T^0{}_0$ denotes the energy density, $p=-T^i{}_j\,\delta^j{}_i/3$ the pressure, and $u_\mu$ is the four-velocity of the fluid, a time-like unit vector. Recalling \eqref{EMT_1f_genspacetime} and using \eqref{flat_RW} we get
 \begin{equation}
\rho=\frac{f(g)}{b^2\sqrt{g}}\left[1-F(g)\right](\phi')^2,
\label{rho1f}
\end{equation}
\begin{equation}
p=\frac{f(g)}{b^2\sqrt{g}}\,F(g)\,(\phi^\prime)^2,
\label{p1f}
 \end{equation}
 where we have considered a homogeneous field $\phi=\phi(\tau)$. 
 It is straightforward to see from equations \eqref{rho1f} and \eqref{p1f} that 
 \begin{equation}
     w_\phi:=\frac{p}{\rho}=\frac{F(g)}{1-F(g)};
     \label{wgen1kin}
 \end{equation}
 which will generally depend on $\tau$ and, thus, the equation of state parameter $w_\phi$ will generally evolve throughout time. One particular case of interest takes place when the coupling function is a power-law, i.e., $f(g)=k g^\alpha$, where $k$ and $\alpha$ are constants. In this case we obtain for $w_\phi$ the following result:
 \begin{equation}
    w_\phi=\frac{\alpha}{1-\alpha}=\mathrm{const.}
     \label{w_1kinpowerlaw}
 \end{equation}
 Notice how this requires $\alpha<1$ in order for the weak energy condition to be satisfied \cite{Jaramillo-Garrido:2023cor}. In addition, the zeroth component of the EMT conservation equation $\nabla_\nu T^{\mu\nu}=0$ yields the usual result \cite{Maroto:2023toq}:
 \begin{equation}
    {\rho'}+3\frac{{a'}}{a}(\rho+p)=0.
    \label{0comp_consEMT}
 \end{equation}

On the other hand, the equation for the $G^{00}$ component of the Einstein tensor yields \cite{Alvarez:2007nn}
 \begin{equation}
     \left(\frac{a'}{a}\right)^2=\frac{8\pi G}{3}\rho b^2,
     \label{Friedmann}
 \end{equation}
 which is the usual Friedmann equation in time $\tau$. Notice that it recovers its original form under the coordinate transformation $\dd t= b(\tau)\,\dd\tau$, where $t$ is the cosmological time. We will denote $'=\dd/\dd\tau$ and $\cdot=\dd/\dd t$.
 
 Finally, let us write the equation of motion of $\phi(\tau)$ in this space-time  \eqref{eq: complete EoM for psi}:
 \begin{equation}
    {\phi''}(\tau)+{\phi'}(\tau)\frac{{L'}(\tau)}{L(\tau)}=0,
     \label{EOM_1f}
 \end{equation}
 where $L(\tau)\equiv f(g(\tau))/b^2(\tau)$. 
This equation of motion implies that 
 \begin{equation}\label{C}
{\phi'}(\tau)=\frac{C_\phi}{L(\tau)},
 \end{equation}
 with $C_\phi$ a constant parameter. Substituting \eqref{C} in equations \eqref{rho1f} and \eqref{p1f}; factoring out $\rho+p$ in equation \eqref{0comp_consEMT} and recalling $g=b^2a^6$, the conservation law \eqref{0comp_consEMT} reads 
 \begin{equation}
    \frac{\dd}{\dd\tau}\ln(a^6)=g'(\tau)\frac{\dd}{\dd g}\left[\ln\left((1-2F(g))\frac{g}{f(g)}\right)\right],
     \label{intermediate_step_1kin}
 \end{equation}
 which provides the geometrical constraint that allows the conservation law \eqref{0comp_consEMT} to be satisfied. This is \cite{Maroto:2023toq}:
 \begin{equation}
    \frac{g}{f(g)}\left(1-2F(g)\right)=C_ga^6,
     \label{cond1kin}
 \end{equation}
 where $C_g$ is a constant. This geometrical constrain on the metric determinant $g$ allows us to obtain the relation between $b$ and $a$ for any given coupling. For instance, if $f(g)\propto g^\alpha$, equation \eqref{cond1kin} implies that 
  \begin{equation}\label{b(a)}
 b\propto a^{3\alpha/(1-\alpha)}.
 \end{equation}
 Notice that only when we take $\alpha=1/2$ (Diff limit), 
 we recover the standard stiff-fluid behaviour $\rho(a)\propto a^{-6}$ of a kinetically dominated scalar field \cite{Gouttenoire:2021jhk,Dutta:2010cu}. In conclusion, TDiff symmetry allows for a much wider phenomenology for simple kinetically driven scalar fields.
\section{Shift-symmetric multi-field TDiff models}\label{secIII}
In this section we will extend the previous results to the case of two free shift-symmetric TDiff homogeneous scalar fields in the matter action with different coupling functions. Since both fields will be kinetically driven, our action will read
\begin{equation}
    \begin{split}
    S_\mathrm{mat}=\int\dd^4x\frac{1}{2}\sum_{i=1}^2\left(f_i(g)\,\partial_\mu\phi_i\partial^\mu\phi_i\right),
    \end{split}
    \label{action2}
\end{equation}
where the respective coupling functions $f_i(g)$ are taken to be positive to avoid the explicit introduction of ghosts in our model \cite{Rubakov:2014jja}. Notice that we did not consider an interaction potential between both fields. As we will see, the energy exchange and the rich phenomenology will arise from geometrical constrains coming from the conservation of the total EMT, since the individual EMTs of each field will not be conserved as a consequence of the symmetry breaking. In fact, since our fields are free, the total EMT will simply be the sum of the individual EMTs of each field:
\begin{align}
    T_{\mu\nu}&=T^{(1)}_{\mu\nu}+T^{(2)}_{\mu\nu}\nonumber \\
    &= (\rho_1+p_1)\,u_\mu u_\nu-p_1\,g_{\mu\nu}+(\rho_2+p_2)\,u_\mu u_\nu-p_2\,g_{\mu\nu}
    \label{total_EMT}
\end{align}
For homogeneous fields in a Robertson-Walker background both fields share a common velocity $u^\mu$ and
\begin{align}
\rho_i=\frac{f_i(g)}{b^2\sqrt{g}}\left[1-F_i(g)\right](\phi_i')^2, \; i=1,2
\label{rhof}
\end{align}
\begin{equation}
p_i=\frac{f_i(g)}{b^2\sqrt{g}}\,F_i(g)\,(\phi^\prime_i)^2, \; i=1,2
\label{pf}
 \end{equation}
  where very much as in the single-field case, we have defined $F_i(g):=\dd\ln f_i(g)/\dd\ln g$, 
so that the corresponding equations of state read
\begin{equation}
     w_i:=\frac{p_i}{\rho_i}=\frac{F_i(g)}{1-F_i(g)},\; i=1,2
     \label{wgenkin}
 \end{equation}

The conservation of the total energy-momentum tensor implies
\begin{align}
    \nabla_\mu T^{\mu\nu}=  \nabla_\mu T^{(1)\mu\nu}+ \nabla_\mu T^{(2)\mu\nu}=0,
\end{align}
which in the Robertson-Walker background reads
\begin{equation}
    {\rho_1'}+3\frac{{a'}}{a}(\rho_1+p_1)+{\rho_2'}+3\frac{{a'}}{a}(\rho_2+p_2)=0.
     \label{tot_rho_cons}
 \end{equation}
Notice that the previous expression does not imply the energy conservation for individual fields, but in general we will have
\begin{align}
    {\rho_1'}&+3\frac{{a'}}{a}(\rho_1+p_1)=Q, \\
     {\rho_2'}&+3\frac{{a'}}{a}(\rho_2+p_2)=-Q, 
     \label{Q_exp}
 \end{align}
 where $Q$ is commonly referred to as the \textit{interacting kernel} in the literature \cite{Bertolami:2012xn}. 
 
 On the other hand, the fields equations of motion read
 \begin{align}
     {\phi_i''}(\tau)+{\phi_i'}(\tau)\frac{{L_i'}(\tau)}{L_i(\tau)}=0, \; i=1,2
 \end{align}
 with $L_i(\tau)\equiv f_i(g(\tau))/b^2(\tau)$, so that very much as in the single-field case, we can write
\begin{equation}
{\phi_i'}(\tau)=\frac{C_{\phi_i}}{L_i(\tau)} \label{phi_p} , \; i=1,2
\end{equation}
with $C_{\phi_i}$ constants. 

Substituting these expressions  into the respective pressures and energy densities \eqref{rhof} and \eqref{pf}, recalling the conservation equation \eqref{tot_rho_cons} and proceeding analogously to the single-field case, calculations yield the following geometrical constrain:
\begin{equation}
    C_{\phi_1}^2\frac{g|2F_1-1|}{f_1}+C_{\phi_2}^2\frac{g|2F_2-1|}{f_2}=C_ga^6,
    \label{constrain2kingen}
\end{equation}

In the case in which the coupling functions are simple power laws
\begin{align}
f_i(g)=\lambda_{i}g^{\alpha_{i}},\hspace{3mm}{i}=1,2;
\end{align}
with $\lambda_i,\alpha_{i}$ constants, the conservation equation \eqref{constrain2kingen} implies 
\begin{equation}
    C_1g^{1-\alpha_1}|2\alpha_1-1|+C_2g^{1-\alpha_2}|2\alpha_2-1|=C_ga^6;
    \label{constrain2kinpowerlaw}
\end{equation}
where $C_1=C_{\phi_1}^2/\lambda_1$ and $C_2=C_{\phi_2}^2/\lambda_2$. This is a very illuminating result, since as we observe from equation \eqref{constrain2kinpowerlaw} it does not require the individual EMT conservation of each field and thus it will involve a geometrical-like interaction between the two components caused by the symmetry breaking. Unlike the single-field case, an explicit solution of this equation cannot be obtained even for simple power-law functions.

Lastly, here we include the expression for the energy density in the power-law coupling case, which will be of valuable use throughout the rest of the work:
\begin{equation}
    \rho_i(a,b)=C_{i}(1-\alpha_i)\frac{b^{1-2\alpha_i}}{a^{6\alpha_i+3}}, \; i=1,2
    \label{rhokinpowerlaw}
\end{equation}
which is straightforwardly obtained from equation \eqref{rhof} using \eqref{phi_p}. Notice how the effective interactions will be reflected on the particular form of $b(a)$ obtained through the EMT conservation law \eqref{constrain2kinpowerlaw}.

 \subsection{Approximate results: single-field domination}
\begin{figure}[h]
    \centering
    \includegraphics[scale=0.65]{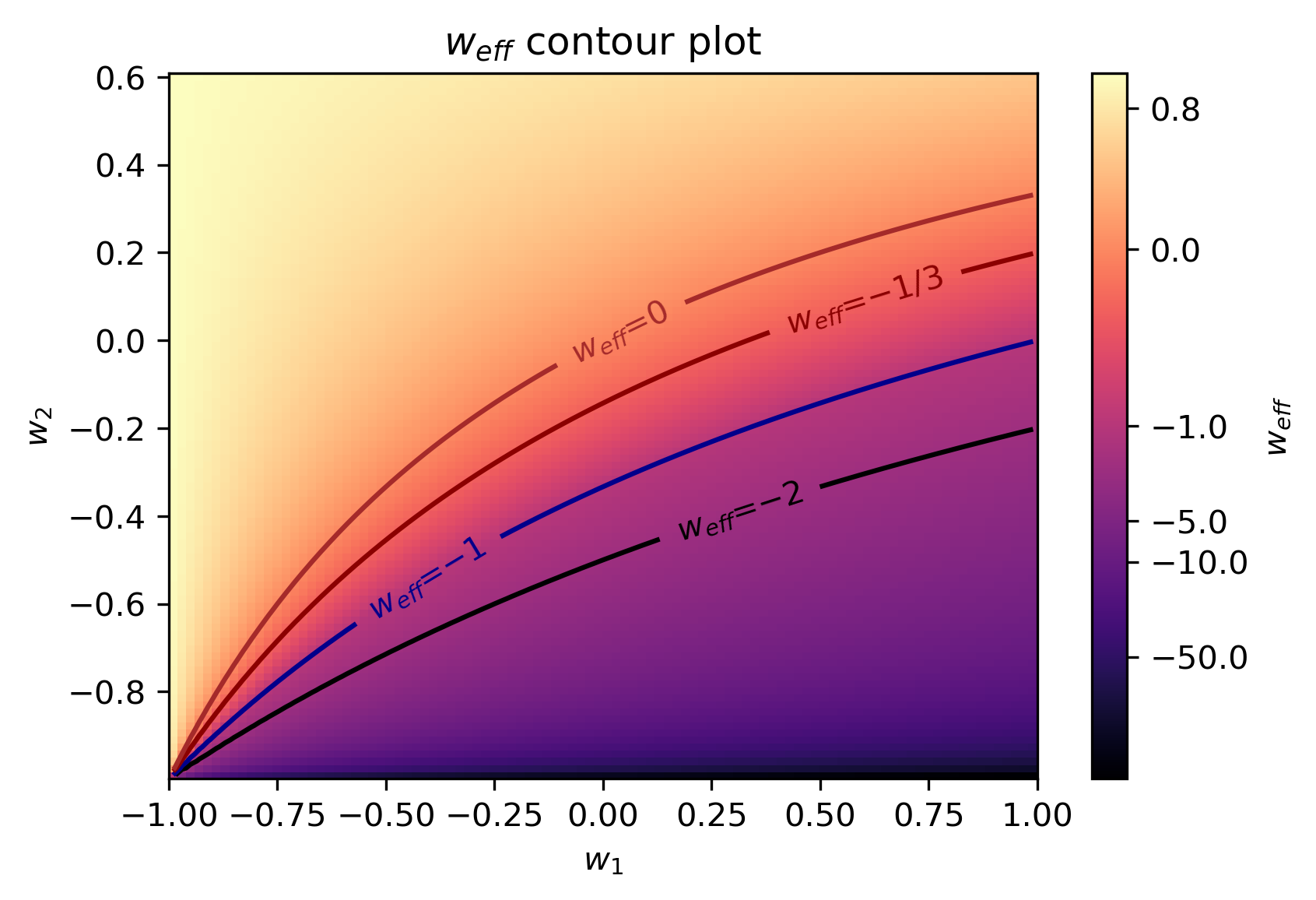}
    \caption{Effective equation of state parameter $w_\mathrm{eff}$ of the subdominant field $\phi_2$ under $\phi_1$ domination in terms of the individual equation of state parameters $w_1$ and $w_2$.}
    \label{2kin_contours}
\end{figure}

 Let us first consider the case in which one of the fields, for example $\phi_1$, dominates over the other, $\phi_2$. We can thus neglect the contribution of $\phi_2$ in \eqref{constrain2kinpowerlaw}, so
 \begin{equation}
    C_1g^{1-\alpha_1}|2\alpha_1-1|\simeq C_ga^6,
     \label{intermediate_step_2kin}
 \end{equation}
 which can be solved as
\begin{align}
 b\propto a^{3w_1} \label{ba1}
 \end{align}
 where 
  \begin{align}
 w_1=\frac{\alpha_1}{1-\alpha_1},\\
 w_2=\frac{\alpha_2}{1-\alpha_2}, \label{w2}
 \end{align}
 as we can see from \eqref{wgenkin}. This is the same geometrical constrain one would obtain if $\phi_1$ was the only field. Notice that this is just an approximation that provides the leading order of $b(a)$, but it gives us valuable information concerning the evolution of the energy densities in the different domination regimes. Recalling \eqref{rhokinpowerlaw} and using \eqref{ba1} yields
 \begin{equation}
     \rho_1(a)\propto a^{-3(1+w_1)}\hspace{2mm}\quad{\rm and}\quad\rho_2(a)\propto a^{-3(1+w_\mathrm{eff})},
     \label{rhoevolution2kin}
 \end{equation}
 and thus $\phi_1$ decays as expected from its equation of state, but the subdominant field $\phi_2$ will exhibit a decay as if it were a perfect fluid with constant equation of state parameter $w_\mathrm{eff}\neq w_2$, where
 \begin{equation}
     w_\mathrm{eff}=\frac{2w_2-w_1+w_1w_2}{1+w_2}.
     \label{weff}
 \end{equation}
 It is worth noting that the individual equation of state parameters $w_i$ will then depict the asymptotic decay behavior of each component when it is dominant. Fig.\ref{2kin_contours} summarizes the wide range of phenomenological possibilities for the subdominant component.

This result happens to be physically illuminating with regards to cosmological contexts. As we can see above, the induced interactions between perfect TDiff fluids with different equation of state parameters allow for a wide range of possible evolutions for the subdominant component. In particular,  all of the possible dark energy behaviors are plausible for the subdominant field, including phantom dark energy \cite{Ludwick:2017tox} ($w_\mathrm{eff}<-1$, where its energy density increases over time) and quintessence. We emphasize that these behaviours can be obtained without the addition of non-canonical kinetic terms \cite{Armendariz-Picon:2000ulo}, they are a result of breaking the Diff symmetry down to TDiff. Interestingly, although $w_i<-1$ is not allowed for each individual field, in accordance to the weak energy condition \cite{Jaramillo-Garrido:2023cor}, the dominance regimes allow for subdominant phantom behavior without violating the energy conditions. As a result, this provides a vast range of possibilities to describe an interacting dark matter-dark energy sector ($w_1=0$, $w_2<-1/3$) with an evolving dark energy decay given by a function $w_\mathrm{eff}(a)$ stemming from the broken Diff invariance, exhibiting phantom decay at early times during the matter epoch. This will allow for phantom-crossing, as it will later be discussed.

\subsection{Energy exchange}
We will now analyze the exchange of energy between the fields induced by the effective interaction, and its evolution through the several field domination regimes by studying the interacting kernel $Q$. Let us consider two kinetically-driven scalar fields $\phi_1$ and $\phi_2$, with constant equation of state parameters $w_1$ and $w_2$, respectively. Let us also assume that $\phi_1$ dominates over $\phi_2$. 
Using \eqref{rhokinpowerlaw} on equation \eqref{Q_exp} and recalling \eqref{weff} we obtain the following expression
\begin{equation}
    Q=3C_2(1-\alpha_2)\frac{{a'}}{a}a^{-3(1+w_\mathrm{eff})}(w_\mathrm{eff}-w_2).
\end{equation}
which can be rewritten as
\begin{equation}
    Q=3\rho_2\mathcal{H}\frac{(w_2-w_1)(1-w_2)}{1+w_2}. \label{Q_p}
\end{equation}
where $\mathcal{H}=a'/a$ denotes the Hubble parameter in time coordinate $\tau$. It is worth mentioning that, according to \eqref{Q_p}, we will not be able to recover the $\Lambda$CDM limit in this model, as when one of the fields starts to approximately behave like a cosmological constant in its asymptotic domination regime ($w_i\rightarrow -1$), $Q$ will diverge and both TDiff components will thus be strongly coupled. This behaviour is linked to the shift-symmetric nature of the fields, which do not have potential terms.

\begin{figure*}[t]
      \centering
      \subfloat{\includegraphics[width=.37\textwidth]{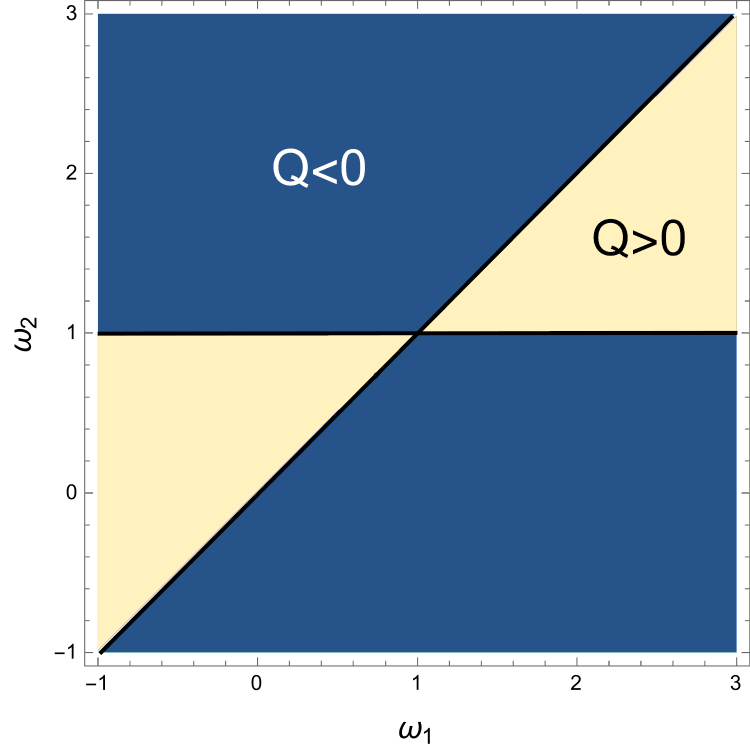}}
      \qquad \qquad
      \subfloat{\includegraphics[width=.37\textwidth]{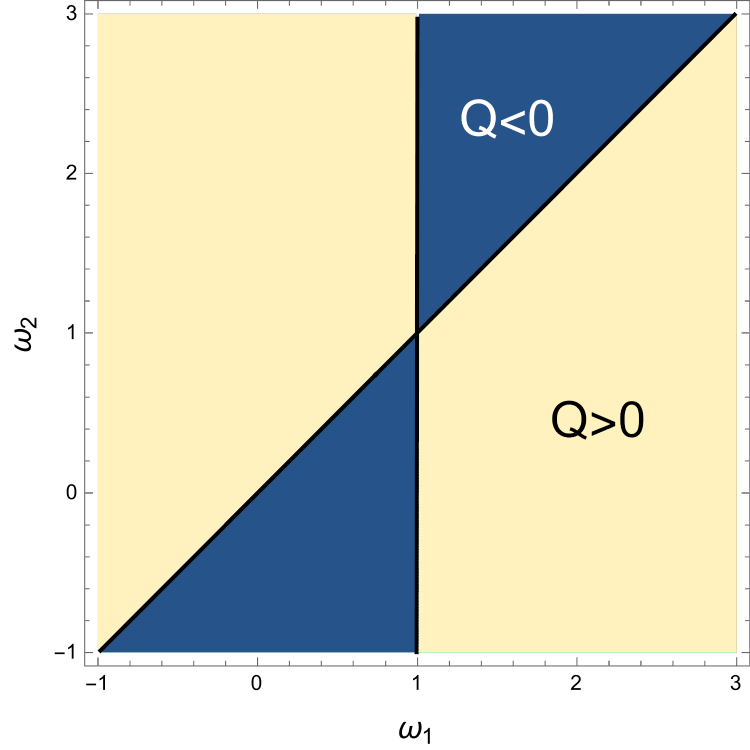}}
      \qquad
  \caption{Sign of the interaction kernel $Q$ when the $\phi_1$ fluid dominates (left) and for $\phi_2$ domination (right)}
  \label{signQ}
\end{figure*}

Let us now study the sign of $Q$ during the single-field domination regimes. Firstly, we observe from \eqref{Q_p} that when $\phi_1$ dominates,  $Q$ has two zeros, those being at $w_\mathrm{eff}=w_2$, i.e., $w_2=w_1$ and $w_2=1$. On the other hand, the analysis in the $\phi_2$ domination regime is fully akin to the previous one, but we have to perform the change $w_1\mapsto w_2$ and change the sign of $Q$ (remember we defined $Q$ with respect to the conservation law for $\phi_1$). We show in Fig. \ref{signQ} the sign of the interaction kernel in both cases ($\phi_1$ and $\phi_2$ domination). 

In light of this analysis, we distinguish three scenarios. In the first case, in which both equation of state parameters are smaller than one ($w_1$, $w_2<1$), the sign of $Q$ does not change between both domination regimes and thus the direction of the energy exchange will not be altered over time. More clearly, if we assume $w_1>w_2$ we see from Fig. \ref{signQ} that when $\phi_1$ dominates $Q<0$ and $\phi_1$ loses energy in favor of $\phi_2$, with the same happening as well when $\phi_2$ is dominant. The same reasoning can be applied to the case in which $w_1<w_2$ (although in this case $Q>0$), allowing us to conclude that in this case it is the field with the greater equation of state parameter who always loses energy.

Secondly, we also have the case in which both fields are beyond stiff fluids ($w_1$, $w_2>1$). We can immediately check (see Fig. \ref{signQ}), similarly to how we proceeded in the previous case, that the direction of the energy exchange will not change during the interaction and it will always be the field with the larger equation of state parameter which gains energy from the other component.

Lastly, there is the case in which one of the fields is beyond a stiff fluid and the other is not ($w_1>1$, $w_2<1$ and vice versa). As opposed to the previous scenarios, we see from Fig. \ref{signQ} that the direction of the energy flux changes between both domination regimes. For instance, if $w_1>1$ and $w_2<1$, $Q$ will be smaller than zero under $\phi_1$ domination and thus $\phi_2$ will be gaining energy from $\phi_1$. However, when $\phi_2$ is dominant, since $w_1>1$ we can see that $Q>0$ and thus it is $\phi_1$ which gains energy from $\phi_2$ now (the analysis is analogous if $w_1<1$, $w_2>1$).

Regarding the potential applications for the description of the dark sector, notice that in \eqref{weff} $w_\mathrm{eff}=-1/3$ when $w_2=(-1+3w_1)/(7+3w_1)\equiv\mathcal{A}<w_1$. This separates the region of $w_2$ values in which the subdominant field, taken to be $\phi_2$ for this example, starts decaying as dark energy. Similarly, $w_\mathrm{eff}=-1$ occurs at $w_2=(-1+w_1)/(3+w_1)\equiv\mathcal{B}$ and it corresponds to the phantom behavior boundary for $\phi_2$. Hence, if $w_1<1$ we can see that if $w_2\in\left(-1,\mathcal{B}\right)$ the subdominant component will exhibit phantom dark energy behavior and the dominant field will lose energy in favor of this; and if $w_2\in(\mathcal{B},\mathcal{A})$ it will also gain energy from the dominant component $\phi_1$, but not enough to display phantom nature.

More specifically, if we consider a dark sector model consisting of dark matter (DM) with $w_1=0$ and dark energy (DE), with $w_2<-1/3$, we can observe from \eqref{weff} that DE will always be phantom during the matter domination epoch due to the energy flux from DM ($Q<0$). The energy exchange will occur in the same direction when DE dominates, although it will now not be enough to keep the phantom behavior, and the DE decay will gradually transition to resemble its asymptotic value for the equation of state parameter $w_2$. On the other hand, DM will slowly start to exhibit a different decay than the typical $a^{-3}$ as DE becomes more dominant.

Lastly, before we go on with our analysis let us briefly comment about the existence of tracking solutions in this model. Recalling \eqref{rhoevolution2kin} we see that the condition that must be satisfied in order for both fields to exhibit the same decay would be
\begin{equation}
    -3(1+w_1)=\frac{3w_1-9w_2-3w_1w_2-3}{1+w_2},
    \label{tracking}
\end{equation}
which cannot be accomplished unless we are in the trivial case in which both components are indeed the same, i.e., $w_1=w_2$, and there would be no interaction. Thus, there will not be tracking solutions in this particular TDiff model.

\subsection{Analytical model}
Solving the general constrain \eqref{constrain2kinpowerlaw} is not a simple task, and it usually requires numerical treatment. However, there is a particular dark sector model for which equation \eqref{constrain2kinpowerlaw} can be analytically solved, consisting of DM with $w_1=0$ $(\alpha_1=0)$ and DE with $w_2=-1/2$ $(\alpha_2=-1)$. Despite not being the best fitting model, as we will later see, being analytical provides us with a wide insight to further understand the physics behind shift-symmetric multi-field TDiff models. The constrain \eqref{constrain2kinpowerlaw} then reads
\begin{equation}
    C_1g+3C_2g^2=C_ga^6,
    \label{cons2kin_analytical}
\end{equation}
which is quadratic in $g$ and can be easily solved as 
\begin{equation}
    g=-\frac{C_1}{6C_2}+\frac{\sqrt{C_1^2+12C_2C_ga^6}}{6C_2},
    \label{gsol_analytical}
\end{equation}
where we have taken into account that $C_i=C_{\phi_i}^2/\lambda_i$ should be positive to avoid ghosts instabilities, so that only the positive-root solution of equation \eqref{cons2kin_analytical} is physically sensible.  
This solution allows us to explicitly obtain the relation $b(a)$:
\begin{equation}
    b(a)=\sqrt{\frac{C_1}{6C_2}}\left[a^{-6}\left(\sqrt{1+\frac{12C_2C_g}{C_1^2}a^6}-1\right)\right]^{1/2},
    \label{b(a)_analytical}
\end{equation}
valid for all values of $a$. As we will later see, the remote past $a\ll1$ will correspond to the matter era, and in the distant future $a\gg1$ DE will be dominant, as expected.

For $a\ll1$, expanding \eqref{b(a)_analytical} in powers of $a$ yields
\begin{equation}
    b(a)\eval_{a\ll1}\simeq\sqrt{C_g}{C_1}\left(1-\frac{3}{2}\frac{C_2C_g}{C_1^2}a^6\right),
    \label{bsmalla}
\end{equation}
from which we can obtain the respective energy densities:
\begin{equation}
    \rho_1(a)\simeq\sqrt{C_1C_g}\left(a^{-3}-\frac{3C_2C_g}{2C_1^2}a^3\right),
    \label{rho1smalla}
\end{equation}
\begin{equation}
    \rho_2(a)\simeq2\left(\frac{C_g}{C_1}\right)^{3/2}C_2a^3.
    \label{rho2smalla}
\end{equation}
Notice how the DM ($\rho_1$) decay is governed by the $a^{-3}$ term, which corresponds to the expected behavior according to $w_1=0$. Consequently, DM is dominant at early times. Besides, DE ($\rho_2$) evolves with $a^3$, exhibiting the phantom nature we previously discussed (in particular, $w_\mathrm{eff}=-2$) as a result of it gaining energy from DM. This can be illustrated writing the conservation equations for each component in terms of the energy density of the other, which read:
\begin{equation}
    {\rho_1'}+3\frac{{a'}}{a}\rho_1\simeq-\frac{9}{2}\mathcal{H}\rho_2,
    \label{cons1smalla}
\end{equation}
\begin{equation}
    {\rho_2'}+3\frac{{a'}}{a}(\rho_2+p_2)\simeq+9C_g^2\frac{C_2}{C_1}\mathcal{H}\frac{1}{\rho_1};
    \label{cons2smalla}
\end{equation}
where the phantom nature is exposed in \eqref{cons2smalla} as a result of $\rho_1$ appearing in the denominator.

On the other hand, for $a\gg1$, expanding \eqref{b(a)_analytical} yields
\begin{equation}
    b(a)\eval_{a\gg1}\simeq\sqrt{\frac{C_1}{6C_2}}\left(\sqrt{A}a^{-3/2}-\frac{1}{\sqrt{2}A}a^{-9/2}\right),
    \label{bgreata}
\end{equation}
with $A\equiv\sqrt{12C_2C_g/C_1^2}$. The energy densities thus read
\begin{equation}
    \rho_1(a)\simeq\frac{C_1^{3/2}}{\sqrt{6C_2}}\left(\sqrt{A}a^{-9/2}-\frac{1}{2\sqrt{A}}a^{-15/2}\right),
    \label{rho1greata}
\end{equation}
\begin{equation}
    \rho_2(a)\simeq\frac{C_1^{3/2}}{6^{3/2}C_2^{1/2}}(-3a^{-9/2}\sqrt{A}+2a^{-3/2}A^{3/2}).    
    \label{rho2greata}
\end{equation}
The leading order for large values of $a$ in $\rho_2$ indicates that now our DE will decay as expected from its equation of state ($w_2=-1/2$), and DM decays faster than $a^{-3}$. This implies that at later times it is DE who becomes dominant, and from the reasoning of the previous subsection, we can see that DM is still giving energy to DE, but not enough to keep the phantom behavior as DM starts becoming subdominant. One could write analogous expressions to \eqref{cons1smalla} and \eqref{cons2smalla}, but they are not as physically enlightening due to the lack of phantom nature under dark energy domination.

We will now write the exact expressions for both energy densities in order to discuss the whole evolution. From the previous analysis we know that both energy densities become equal at a certain time: $a=a_\mathrm{eq}$, $\rho_i=\rho_\mathrm{eq}$. We can thus write $\rho_1$ and $\rho_2$ substituting \eqref{b(a)_analytical} in \eqref{rhokinpowerlaw} and equating them. We obtain
\begin{equation}
    C_1=\frac{5}{2}\rho_\mathrm{eq}^2a_\mathrm{eq}^6C_g^{-1}\hspace{2mm},\hspace{2mm}C_2=\frac{125}{16}\rho_\mathrm{eq}^4a_\mathrm{eq}^6\frac{1}{C_g^3};
    \label{constants_changed}
\end{equation}
which allow us to write down the energy densities in terms of these parameters, which are easier to physically interpret than the integration constants $C_i$. Thus, we have:
\begin{equation}
    \rho_1(z)=\frac{1}{\sqrt{3}}\rho_\mathrm{eq}\left(\frac{1+z}{1+z_\mathrm{eq}}\right)^6\Theta^{1/2}(z),
    \label{rho1_newconst}
\end{equation}
\begin{equation}
    \rho_2(z)=\frac{1}{\sqrt{27}}\rho_\mathrm{eq}\left(\frac{1+z}{1+z_\mathrm{eq}}\right)^6\Theta^{3/2}(z);
    \label{rho2_newconst}
\end{equation}
where we defined $\Theta(z)\equiv\sqrt{1+15[(1+z_\mathrm{eq})/(1+z)]^6}-1$, and where $z=1/a-1$ denotes the redshift.

We will now obtain some physical results and compare this model to $\Lambda$CDM before analyzing the general case. Firstly, recalling \eqref{rhokinpowerlaw} and the conservation law \eqref{0comp_consEMT}, we can parameterize the decay of each component with a function $w_\mathrm{eff,i}(z)$ which satisfies the individual conservation law
\begin{equation}
    {\rho'}_i+3\frac{{a'}}{a}[1+w_\mathrm{eff,i}(z)]\,\rho_i=0;
    \label{cons_eff}
\end{equation}
which yields the following result when recalling \eqref{rhokinpowerlaw}:
\begin{equation}
    w_\mathrm{eff,i}(z)=-\frac{1}{3}\left[-\frac{1+z}{b(z)}(1-2\alpha_i)\frac{\dd b}{\dd z}-6\alpha_i-3\right]-1,
    \label{weff(z)}
\end{equation}
where $\alpha_i$ denote the exponents of the coupling functions of each component. This recovers the previously studied constant results when considering the respective field domination regimes. The functions $w_\mathrm{eff,i}(z)$ can easily be simplified for the analytical case since we know $b(z)$ explicitly. 

\begin{figure}[h]
    \centering
    \includegraphics[scale=0.60]{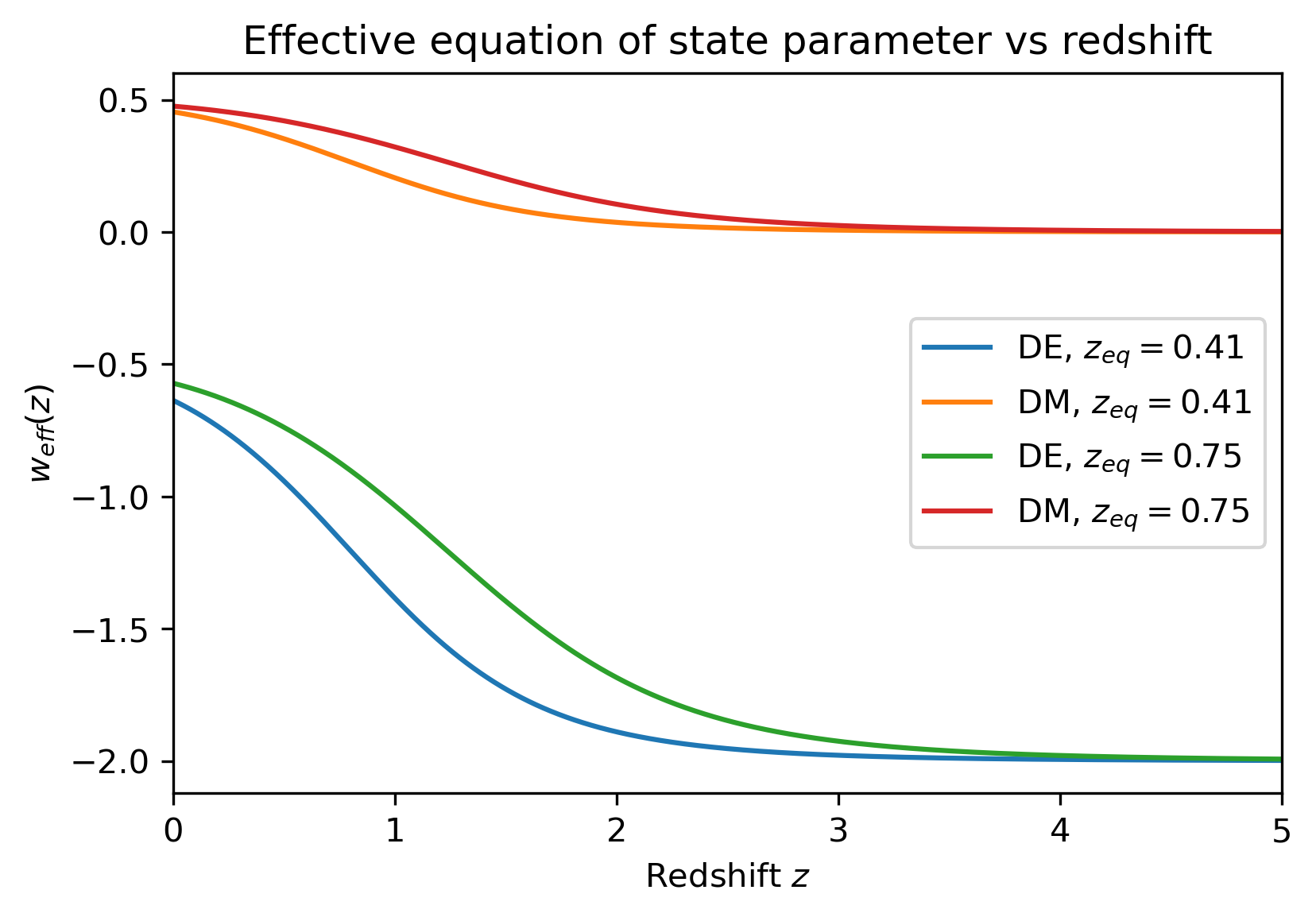}
    \caption{$w_\mathrm{eff}(z)$ for DM and DE for various $z_\mathrm{eq}$. Dark energy transitions from being phantom during the matter era to decaying as usual dark energy with equation of state parameter $w_2$, with there being phantom crossing at recent times. Dark matter starts to decay faster than expected from $w_1=0$ as dark energy starts dominating.}
    \label{weff_vec_analytical}
\end{figure}

Using the exact expression for $b(z)$ \eqref{b(a)_analytical} yields the result in Fig.\ref{weff_vec_analytical}, for two different values of the free parameter $z_\mathrm{eq}$.
Notice how in reality we only have $z_\mathrm{eq}$ as our free parameter, since the cosmic sum rule $(\rho_1+\rho_2)\eval_{t=t_0}=(1-\Omega_\mathrm{B})\rho_\mathrm{c}$ must be satisfied and, thus, it enforces an extra relation between the parameters that allows to remove the dependence on $\rho_\mathrm{eq}$. As a reminder, $\Omega_\mathrm{B}$ depicts the baryonic matter component of the universe, $\rho_\mathrm{c}$ is the critical density and we are ignoring the radiation component at late times. It is worth noting that in this work we assumed that only the dark sector breaks the Diff invariance, hence we will treat baryons as ordinary Diff matter. As we can see from Fig.\ref{weff_vec_analytical}, the DE decay behavior starts being phantom-like at early times, as expected, and then evolves in time until it reaches the asymptotic value reflected in the equation of state $w_2>-1$ in the future, with there being phantom-crossing near the present. The parameter $z_\mathrm{eq}$ only changes slightly the behavior in the intermediate regimes, without altering the main physical behavior. On the other hand, DM will exhibit its usual $a^{-3}$ decay at early times but it will decay faster when DE starts to dominate. In light of this we can see that TDiff models can provide a very rich phenomenology involving different time evolutions for the dark sector. This could lead to new models for interactions in the dark sector without the introduction of non-canonical terms or ghost instabilities in the action. 

\begin{figure}[h]
    \centering
    \includegraphics[scale=0.60]{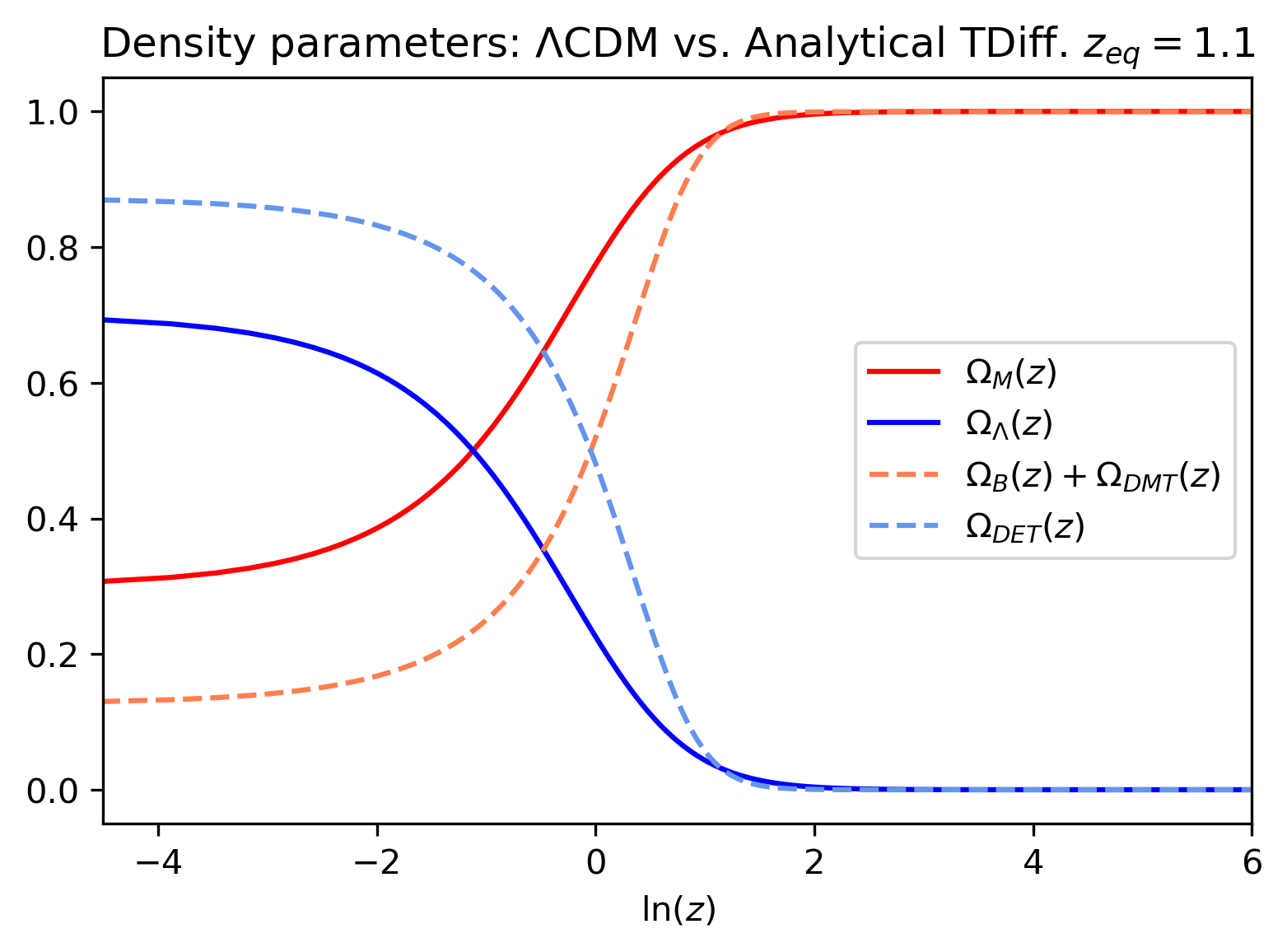}
    \caption{Evolution of the density parameters: $\Lambda$CDM (continuous lines) vs analytical TDiff case (dashed lines), for $z_\mathrm{eq}=1.10$. In shift-symmetric TDiff models dark energy would be more dominant as a consequence of it being phantom at early times, gaining energy from dark matter.}
    \label{Omegas_analytical}
\end{figure}

Lastly, we will analyze this model from the perspective of the density parameters to further understand shift-symmetric TDiff dark sector models. We will denote the density parameters for DM and DE, respectively, as $\Omega_\mathrm{DMT}$ and $\Omega_\mathrm{DET}$. We will also use the standard notation for the $\Lambda$CDM parameters: $\Omega_\mathrm{M}=\Omega_\mathrm{DM}+\Omega_\mathrm{B}$ for matter and $\Omega_\Lambda$ for the cosmological constant. Recalling the Friedmann equation \eqref{Friedmann} and using cosmological time $\dd t=b(\tau)\dd\tau$ yields
\begin{equation}
    H^2=\frac{8\pi G}{3}(\rho_\mathrm{B}+\rho_1+\rho_2).
    \label{Friedmann_analytical}
\end{equation}
Multiplying and dividing this expression by the Hubble parameter at $t=t_0$ (today) $H_0^2$, and recalling that $\rho_\mathrm{c}=8\pi G/(3H_0^2)$ it is straightforward to obtain \correct{the respective abundances in terms of the redshift, $\Omega_i(z)=8\pi G\rho_i(z)/[3H^2(z)]$:}
\begin{equation}
    \Omega_\mathrm{DMT}(z)=\frac{1}{\sqrt{3}}\frac{\rho_\mathrm{eq}}{\rho_\mathrm{c}}\left(\frac{1+z}{1+z_\mathrm{eq}}\right)^6\frac{\Theta^{1/2}(z)}{E^2(z)},
    \label{OmegaDM_analytical}
\end{equation}
\begin{equation}
    \Omega_\mathrm{DET}(z)=\frac{1}{3\sqrt{3}}\frac{\rho_\mathrm{eq}}{\rho_\mathrm{c}}\left(\frac{1+z}{1+z_\mathrm{eq}}\right)^6\frac{\Theta^{3/2}(z)}{E^2(z)},
    \label{OmegaDE_analytical}
\end{equation}
where we defined
\begin{equation}
    E^2(z)\equiv\Omega_\mathrm{B}(1+z)^3+\frac{\rho_\mathrm{eq}}{\sqrt{3}\rho_\mathrm{c}}\left(\frac{1+z}{1+z_\mathrm{eq}}\right)^6\Theta^{1/2}(z)\left[1+\frac{\Theta(z)}{3}\right].
    \label{E(z)_analytical}
\end{equation}
This allows us to obtain the time evolution for each density parameter. This will obviously depend on the specific value of $z_\mathrm{eq}$, but the general behaviour will be similar. For the sake of simplicity, we included the case $z_\mathrm{eq}=1.1$ in the Fig.\ref{Omegas_analytical} to show qualitative results.
This results in a higher DE abundance and a lower DM one at $t=t_0$ than those from $\Lambda$CDM, which can be interpreted as a consequence of the phantom era during the DM domination regime. As we previously discussed, DM transfers part of its energy to DE, which translates into its phantom behavior and thus contributes to obtaining higher values of $\Omega_\mathrm{DET}$. It is worth mentioning, however, that we shall not directly compare these parameters to those from $\Lambda$CDM, as $\Omega_\mathrm{DMT}$ and $\Omega_\mathrm{DET}$ may not be regarded as \textit{true} DM and DE density parameters, since, as opposed to $\Lambda$CDM, this model presents an interacting dark sector and thus there may be contributions from both components to each parameter. 

\section{A TDiff model for dark sector interactions}\label{secIV}
We will now consider a more general case with $w_1=0$, which could play the role of DM; and arbitrary $w_2<-1/3$, which could play the role of DE. We will then contrast the predictions of this simple model to observations to get a glimpse on the viability of shift-symmetric TDiff models for describing the dark sector. Recalling the geometrical constrain \eqref{constrain2kinpowerlaw} arisen from the conservation of the EMT, using $\alpha_1=0$ and dividing by $C_2$ yields
\begin{equation}
    \lambda(1-\alpha_2)g+g^{1-\alpha_2}|2\alpha_2-1|=a^6\left(\lambda(1-\alpha_2)+|2\alpha_2-1|\right).
    \label{numeric_cons_gen}
\end{equation}
where we have used \begin{equation}
    \frac{\rho_1(t_0)}{\rho_2(t_0)}=\frac{C_1}{C_2}(1-\alpha_2)^{-1}=\frac{\Omega_\mathrm{DMT}}{\Omega_\mathrm{DET}}\equiv\lambda.
    \label{cons_inter_2}
\end{equation}
and normalized\footnote{Notice how we can fix the second condition $g(t_0)=1$ as well, since performing this change will be reflected in the action \eqref{action2} as a global constant $\mathcal{C}$, $\Tilde{S}_\mathrm{mat}=\mathcal{C}S_\mathrm{mat}$; and therefore both actions will be physically equivalent under a redefinition of the fields embodying this change: $\phi_i\mapsto\Tilde{\phi}_i=\sqrt{\mathcal{C}}\phi_i$.} $a(t_0)=1$, and also $g(t_0)=1$, which leads to $b(t_0)=1$.
Therefore, we only have two free parameters, those being the exponent of the power-law coupling function of the DE component, $\alpha_2$, and $\lambda$. However, we will use another physical parameter instead of $\lambda$ in order to obtain a more direct analysis and an easier comparison to observations. In fact, recalling Friedman equation \eqref{Friedmann_analytical} in cosmological time, using \eqref{rhokinpowerlaw} and noting that $(\rho_1+\rho_2)\eval_{t_0}=(1-\Omega_\mathrm{B})\rho_\mathrm{c}$ as a consequence of the cosmic sum rule yields the following expression for $H^2(z)$:
\begin{equation}
    \begin{split}
    H^2(z)=H_0^2\left[\Omega_\mathrm{B}(1+z)^3+(1-\Omega_\mathrm{B})\left(1+\frac{1}{\lambda}\right)^{-1}b(z)(1+z)^3\right. & \\ \left.+(1-\Omega_\mathrm{B})\frac{1}{\lambda}\left(1+\frac{1}{\lambda}\right)^{-1}b(z)^{1-2\alpha_2}(1+z)^{6\alpha_2+3}\right],
    \end{split}
    \label{H(z)_numeric_general}
\end{equation}
where we neglected radiation, as the purpose of this model is to study the DM and DE domination epochs. Otherwise, we should have included the corresponding $\Omega_\mathrm{R}(1+z)^4$ contribution from radiation, \correct{where we are assuming it is a Diff component.} \footnote{\correct{Hence, its EMT is automatically conserved under solutions of Einstein equations and at early times during the radiation era the model behaves the same way as $\Lambda$CDM, as the TDiff components are negligible at such time. Thus, the bounds imposed by BBN are not modified in this model.}}. Notice that at early times, when the $\phi_1$ fluid dominates over $\phi_2$, we can neglect the last term in \eqref{H(z)_numeric_general}. In addition, 
$b(a)\propto a^{3w_1}$ takes a constant value at early times  $b(z)\simeq b_{early}$ since $w_1=0$. This allows us to define the following effective density parameter for total matter at high redshift
\begin{equation}
    \Omega_\mathrm{M}^\mathrm{eff}\equiv\Omega_\mathrm{B}+(1-\Omega_\mathrm{B})\left(1+\frac{1}{\lambda}\right)^{-1}b_\mathrm{early}.
    \label{OMeff}
\end{equation}
We will use this parameter $\Omega_\mathrm{M}^\mathrm{eff}$ instead of $\lambda$, since both are trivially related through \eqref{OMeff}. Acknowledge that $b_\mathrm{early}$ can be directly computed from the conservation equation \eqref{numeric_cons_gen} taking into consideration that DM dominates at this time and radiation does not contribute to the geometrical constrain, since we are treating it as a Diff component and thus its EMT is automatically conserved. Thus, using \eqref{cons_inter_2} we obtain
\begin{equation}
    b_\mathrm{early}=\sqrt{\frac{\lambda(1-\alpha_2)+|2 \alpha_2-1|}{\lambda(1-\alpha_2)}}.
    \label{b_early}
\end{equation}

If we express $H_0^{-1}$ as $2997.9h^{-1}$ Mpc, with $h$ being the reduced Hubble constant, this will allow us to fit our parameters $( w_2, \Omega_\mathrm{M}^\mathrm{eff})$ to observations and obtain physical predictions for this model. Notice that we are using $w_2$ instead of $\alpha_2$ as the 
model parameter since the are trivially related by \eqref{w2}.
We will consider the baryon density parameter obtained from the abundance of light elements $\Omega_\mathrm{B}h^2=0.02240\pm0.00069$ \cite{ParticleDataGroup:2022pth}, as it is independent of the particular choice for the cosmological model, \correct{and we will marginalize the absolute magnitude $M$, which is equivalent to marginalize $H_0$ since they are degenerated, as we will later explain} (\correct{for the supernovae analysis}) \cite{Planck:2018vyg}. In particular, we developed a code in \textit{Python} that solves the conservation law \eqref{numeric_cons_gen} for any given pair of these two parameters. Hence, we can obtain $b(z)$ and $H(z)$ through \eqref{H(z)_numeric_general}. \correct{We will then regard different data sets (namely Union2 supernovae and CMB) and perform a preliminary numerical likelihood analysis in order to study the validity of the model. We will then present the structure of the analysis in the following subsections.}

\correct{\subsection{Union2 supernovae data set}}
\correct{We will first consider the Union2-database observational data coming from type Ia Supernovae \cite{Scolnic:2021amr,Brout:2022vxf} consisting of 557 data for $0.015<z<1.030$ and compare them to the theoretical distance moduli $\mu(z)$ predictions of our model. We will study the agreement between theory and observations using the $\chi^2$ statistical estimator \cite{Amanullah:2010vv}:
\begin{equation}
    \chi^2_\mathrm{SNIa}=\sum_i\frac{(\mu^\mathrm{obs}(z_i)-\mu^\mathrm{th}(z_i))^2}{E_i^2},
    \label{chi2_1parametro}
\end{equation}
where $E_i$ denotes the error in the $\mu_i$ measurement at redshift $z_i$ and the theoretical distance modulus is given by
\begin{equation}
    \mu^\mathrm{th}(z)=5\log_{10}\left(\frac{d_\mathrm{L}(z)}{1\hspace{1mm}\mathrm{Mpc}}\right)+M=\hat{\mu}(z)+M,
    \label{distance_modulus}
\end{equation}
with $d_\mathrm{L}(z)$ the luminosity distance computed from
\begin{equation}
    d_\mathrm{L}(z)=(1+z)\int_0^z\frac{\dd z}{H(z)},
\end{equation}
for flat spatial sections and $M$ being the absolute magnitude, which we marginalized the following way:
\begin{equation}
    M=\sum_i\left(\frac{1}{\sigma}\frac{\mu^\mathrm{obs}(z_i)-\hat{\mu}(z_i)}{E_i^2}\right).
    \label{M_marg}
\end{equation}
 with $\sigma=\sum_i E_i^{-2}$. \correct{Notice that this is equivalent to marginalizing $H_0$, as it is degenerated with $M$, according to \eqref{distance_modulus}.} Numerical integration will allow us to perform the analysis in the subsequent sections.}\\
 
 \correct{\subsection{CMB data set}
We will also consider the CMB data to study the observational viability of our model. For this purpose, we will be using the two CMB distance priors $R$ (the shift parameter) and $\ell_a$ (the acoustic length) \cite{Pl2015,Alonso-Lopez:2023hkx} instead of the Planck 2018 full likelihood, as our model behaves the same way as $\Lambda$CDM at early times and these parameters allow us to easily assemble all the relevant information. The respective values measured for these parameters for the Planck 2018 TT,TE,TE+lowE+lensing data \cite{Planck:2018vyg} are the following \cite{Alonso-Lopez:2023hkx}:
\begin{align}
    R=1.7497\pm0.0041,
    \label{R_exp} \\
    \ell_a=301.529\pm0.083,
    \label{la_exp}
\end{align}
with the covariance matrix given by 
\begin{equation}
    \mathbf{Cov}= \begin{pmatrix}
6.889\cdot10^{-3} & 1.2090859\cdot10^{-4} \\
1.2090859\cdot10^{-4} & 1.681\cdot10^{-5} \\
\end{pmatrix},
\label{covariance}
\end{equation}
which was obtained using the respective correlation matrix presented in \cite{Alonso-Lopez:2023hkx}. 

On the other hand, the theoretical expressions used to compute the distance priors read
\begin{align}
    R=\sqrt{\Omega_M^\mathrm{eff}H_0^2}(1+z_*)d_\mathrm{A}(z_*), 
    \label{R_theo} \\
    \ell_a=\pi(1+z_*)\frac{d_\mathrm{A}(z_*)}{r_\mathrm{s}},
    \label{la_teo}
\end{align}
where $d_\mathrm{A}(z)$ denotes the angular distance 
\begin{equation}
    d_\mathrm{A}(z)=\frac{1}{1+z}\int_0^z\frac{\dd z'}{H(z')},
\end{equation}
and $r_\mathrm{s}$ denotes the radius of the sound horizon
\begin{equation}
    r_\mathrm{s}=r(z_*)=\int_{z_*}^\infty\dd z'\frac{c_\mathrm{s}(z')}{H(z')},
    \label{rs_teo}
\end{equation}
where $c_\mathrm{s}(z)$ is the speed of sound in the photon-baryon fluid. The quantity $z_*$ present in the formulas above depicts the decoupling redshift, whose value is obtained through the fitting expressions in \cite{smallscale}:
\begin{equation}
z_*=1048(1+0.00124\,\omega_\mathrm{B}^{-0.738})(1+g_1\omega_\mathrm{M}^{g_2}),
\label{z_star}\\
\end{equation}
\begin{align}
g_1=\frac{0.0783\omega_\mathrm{B}^{-0.238}}{1+39.5\omega_\mathrm{B}^{0.763}},
\label{g1} \\
g_2=\frac{0.560}{1+21.1\omega_\mathrm{B}^{1.81}}.
\end{align}
It is worth mentioning that $\omega_\mathrm{B}$ and $\omega_\mathrm{M}$ denote the respective reduced baryonic and matter density parameters, i.e., $\omega_\mathrm{B}=\Omega_\mathrm{B}h^2$ and $\omega_\mathrm{M}=\Omega_\mathrm{M}^\mathrm{eff}h^2$, where we are using $\Omega_\mathrm{M}^\mathrm{eff}$ as all of the previous expressions are meant to be evaluated at high redshift, where $\Omega_\mathrm{M}^\mathrm{eff}$ is a constant in light of \eqref{OMeff} and acts as the usual matter abundance. Similarly, we also must take the radiation term into account in the Hubble rate, that is $\Omega_\mathrm{R}(1+z)^4$, when performing this calculations. 

Finally, we will also study the accordance between our model and these distance prior data using the $\chi^2$ estimator:
\begin{equation}
\chi^2_\mathrm{CMB}=\mathbf{\Delta}^T\cdot\mathbf{Cov}^{-1}\cdot\mathbf{\Delta},
\label{chi2_CMB}
\end{equation}
where $\mathbf{\Delta}$ depicts the vector consisting of the differences between the distance prior data and their theoretical values, which depend on the parameters of our TDiff model We will lastly consider the full $\chi^2$ combined function:
\begin{equation}
    \chi^2=\chi^2_\mathrm{SNIa}+\chi^2_\mathrm{CMB}.
    \label{chi_tot}
\end{equation}

}

\subsection{Two-parameter fit}
We will now analyze our TDiff model to conclude if the actual best TDiff fit is compatible with type Ia supernovae observations \correct{and CMB data}. On the grounds of this, \correct{for the type Ia supernova fit we considered a grid of values of $w_2$ $\in$ (-0.993,-0.50) and $\Omega_\mathrm{M}^\mathrm{eff}$ $\in$ (0.10,0.50) and computed $\chi^2_\mathrm{SNIa}$ numerically from \eqref{chi2_1parametro} for the grid marginalizing $M$ (or, equivalently, $H_0$, as they are degenerated). For the CMB fit we also marginalized $H_0$, but in a numerical way after having computed the full likelihood grid, as we cannot proceed analytically like we did in \eqref{M_marg}. We thus considered a parameter grid consisting of values of $w_2\in(-0.80,-0.42)$, $\Omega_\mathrm{M}^\mathrm{eff}\in(0.23,0.41)$ and $h\in(0.59,0.77)$ and computed $\chi^2_\mathrm{CMB}$ for each case using the distance priors and \eqref{chi2_CMB}. We then marginalized the Hubble constant and obtained the two-parameter CMB likelihood that we will use to compare both data analyses.} In the following analysis we will also include the direct $w$CDM analogue of this two-parameter fit, in which both $\Omega_\mathrm{M}$ and $w$ are fitted (using values for $w$ in $(-1.75,-0.45)$ and $\Omega_\mathrm{M}$ in $(0.05,0.60)$), in order to compare both models. Numerical analysis thus yields the results in Tab.\ref{tab2p}-\ref{tab2ptotal}.

\begin{table}[H]
\centering
\begin{tabular}{|c|c|c|}
\Xhline{1\arrayrulewidth}
       SNIa      & Best fit                                                             & $\chi^2_\mathrm{min}$ \\ 
\Xhline{1\arrayrulewidth}
TDiff        & $w_2=-0.813_{-0.060}^{+0.102}$, $\Omega_\mathrm{M}^\mathrm{eff}=0.387_{-0.078}^{+0.056}$ & 542.16                \\ 
\Xhline{1\arrayrulewidth}
$w$CDM & $w=-1.142^{+0.145}_{-0.184}$, $\Omega_\mathrm{M}=0.346_{-0.083}^{+0.082}$                                & 542.64                \\ 
\Xhline{1\arrayrulewidth}
\end{tabular}
\caption{Two-parameter fit (SNIa): TDiff vs $w$CDM. Both models show similar agreement with observations. The 1-$\sigma$ intervals for each parameter have also been included.}
\label{tab2p}
\end{table}

\begin{table}[H]

\centering
\begin{tabular}{|c|c|c|}
\Xhline{1\arrayrulewidth}
       CMB      & Best fit                                                             & $\chi^2_\mathrm{min}$ \\ 
\Xhline{1\arrayrulewidth}
  TDiff        & $w_2=-0.722^{+0.191}_{-0.058}$, $\Omega_\mathrm{M}^\mathrm{eff}=0.263^{+0.077}_{-0.016}$ & -                \\ 
\Xhline{1\arrayrulewidth}
$w$CDM & $w=-1.278_{-0.042}^{+0.378}$, $\Omega_\mathrm{M}=0.236_{-0.002}^{+0.097}$                                & -                \\ 
\Xhline{1\arrayrulewidth}
\end{tabular}
\caption{Two-parameter fit: TDiff vs $w$CDM (CMB). Both models show agreement with observations. The 1-$\sigma$ intervals for each parameter have also been included. The - in the $\chi^2$ column indicates that $\chi^2$ is zero for the best fit, as expected since we are fitting two parameters using data for two distance priors. The 68\% regions are very asymmetric as a consequence of the non-gaussianity of the probability distributions.}
\label{tab2pCMB}
\end{table}

\begin{table}[H]

\centering
\begin{tabular}{|c|c|c|}
\Xhline{1\arrayrulewidth}
       \text{\small SNIa+CMB}      & Best fit                                                             & $\chi^2_\mathrm{min}$ \\ 
\Xhline{1\arrayrulewidth}
TDiff        & $w_2=-0.703^{+0.026}_{-0.026}$, $\Omega_\mathrm{M}^\mathrm{eff}=0.273^{+0.010}_{-0.010}$ & 557.97                \\ 
\Xhline{1\arrayrulewidth}
$w$CDM & $w=-1.092_{-0.034}^{+0.034}$, $\Omega_\mathrm{M}=0.292_{-0.010}^{+0.010}$                                &      556.63           \\ 
\Xhline{1\arrayrulewidth}
\end{tabular}
\caption{Two-parameter fit: TDiff vs $w$CDM (SNIa+CMB). Both models are in good agreement with observations. The 1-$\sigma$ intervals for each parameter have also been included. There is a difference of less than  1-$\sigma$ between both models.}
\label{tab2ptotal}
\end{table}

\begin{figure}
    \centering
    \includegraphics[scale=0.85]{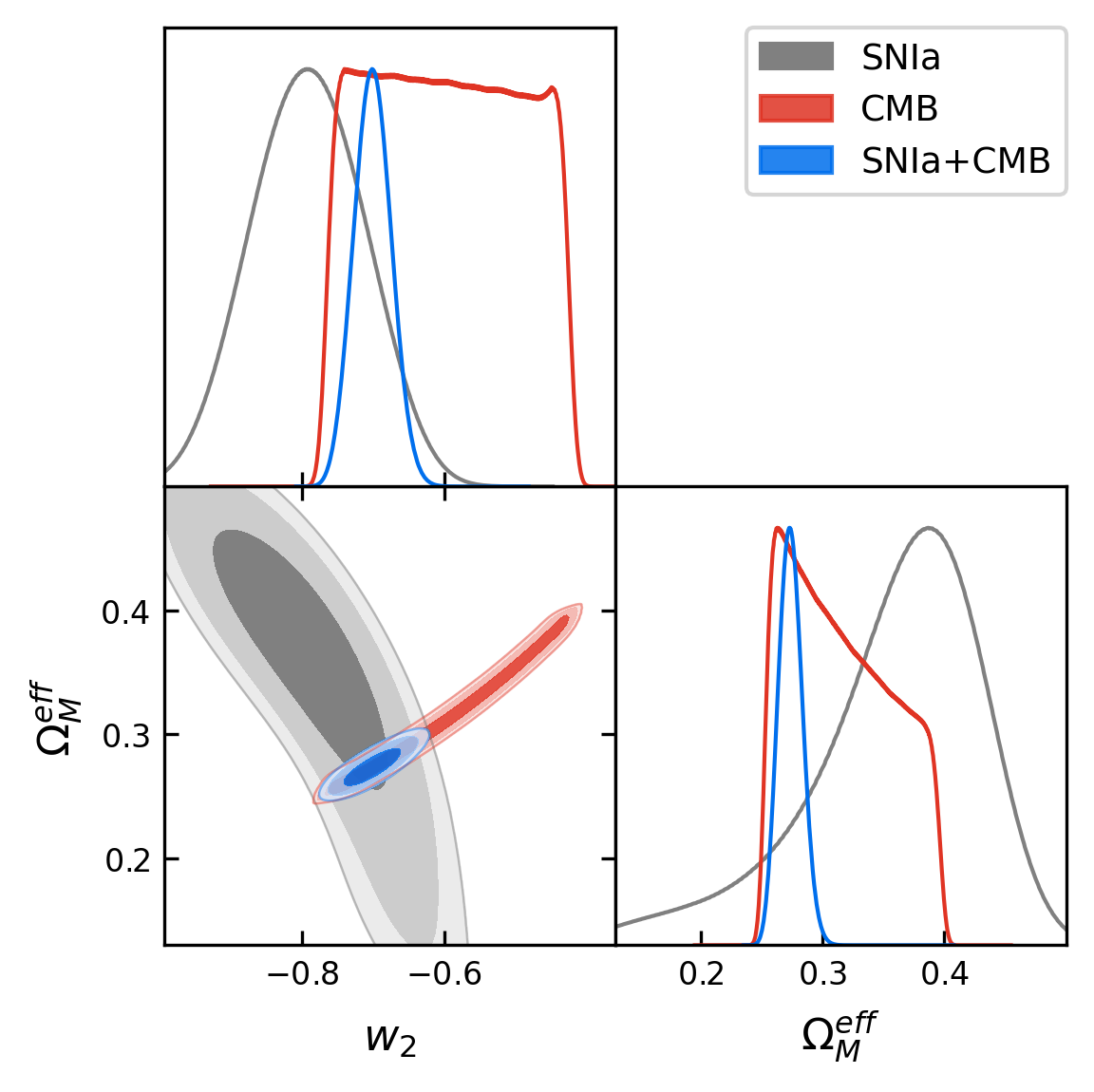}
    \caption{Two-parameter fit: contour plot for $\chi^2$ up to the 3-$\sigma$ region using Union2 and CMB data. The 1-$\sigma$ region for the SNIa and CMB are compatible with each other. Notice that the 68\% contour region differs a bit from the marginalized 1-$\sigma$ intervals, which is a consequence of the non-gaussianity of the distributions.}
    \label{patricio}
\end{figure}

These results indicate that both TDiff and $w$CDM fit well with the observational data, with the \correct{difference between the joint fits for both models being lower than 1-$\sigma$ (although $w$CDM presents a slightly better goodness of the fit)}. Therefore, we will focus on the TDiff case from now on and present the contour plot for both parameters up to the 3-$\sigma$ region in Fig.\ref{patricio}.
 \correct{The results obtained} indicate that TDiff models provide good compatibility when it comes to type Ia supernovae and CMB observations. \correct{It is also worth mentioning that the marginalized distributions for our parameters in the CMB fit are not gaussian and exhibit abrupt decays at the extremes, which can be a consequence of a strong correlation and degeneracy between our parameters (see Fig.\ref{patricio}). Lastly, it is worth recalling that this is an exploratory analysis that enables us to check the observational viability of the model and get some constraints in relation with the cosmological parameters. We are also using an approximate CMB likelihood, and thus the full likelihood using other observational sets and all of the observables should be considered in the future. }

 The respective 1-$\sigma$ intervals for each of the parameters were obtained by marginalizing the joint likelihood
\begin{equation}
    \mathcal{L}(w_2,\Omega_\mathrm{M}^\mathrm{eff})=\mathcal{N}\mathrm{e}^{-\chi^2_\mathrm{s}/2},
    \label{likelihood}
\end{equation}
which can be done for each variable by performing the integration with respect to the other. This yields:
\begin{equation}
    \mathcal{L}^\mathrm{m}(w_2)=\mathcal{N}_1\int\mathrm{e^{-\chi^2_\mathrm{s}/2}}\dd \Omega_\mathrm{M}^\mathrm{eff},\hspace{2mm} \mathcal{L}^\mathrm{m}(\Omega_\mathrm{M}^\mathrm{eff})=\mathcal{N}_2\int\mathrm{e^{-\chi^2_\mathrm{s}/2}}\dd w_2;
    \label{Lmar_beta}
\end{equation}
where $\mathcal{N}_1$ and $\mathcal{N}_2$ are normalization constants (the way of proceeding for the $w$CDM case is fully analogous).  The maximization of these marginalized likelihood distributions $\mathcal{L}^\mathrm{m}$ allowed us to obtain the 1-$\sigma$ regions for both of the parameters, which has been done making use of the GetDist package for Python\footnote{\url{https://getdist.readthedocs.io/}.}. \correct{It is worth mentioning that we can estimate the tension between the SNIa and CMB results for $w_2$ and $\Omega_\mathrm{M}^\mathrm{eff}$ using the following expressions:
\begin{equation}
    \tau_1=\frac{\left|w_2^\mathrm{SNIa}-w_2^\mathrm{CMB}\right|}{\sqrt{\sigma_{w,\mathrm{SNIa}}^2+\sigma_{w,\mathrm{CMB}}^2}},\hspace{2mm}    \tau_2=\frac{\left|\Omega_\mathrm{M}^\mathrm{eff,\mathrm{SNIa}}-\Omega_\mathrm{M}^\mathrm{eff,\mathrm{CMB}}\right|}{\sqrt{\sigma_{\Omega,\mathrm{SNIa}}^2+\sigma_{\Omega,\mathrm{CMB}}^2}},
\end{equation}
where the $\sigma_i$ denote the respective error of the result. Since our distributions are not gaussian, we can get an approximation considering the respective half-length of the respective 68\% intervals. This yields $\tau_1\simeq0.64\sigma$ and $\tau_2\simeq
1.12\sigma$, which are small and indicate that both data sets can be combined for a joint analysis.

Lastly, in order to compare the performance of both our TDiff model and $w$CDM, it will be useful for us to check the DIC coefficient (deviance information criteria), which can be obtained for each model the following way \cite{DIC}:
\begin{equation}
    \mathrm{DIC}=2\Bar{\chi}^2(\mathbf{x})-\chi^2(\Bar{\mathbf{x}}),
    \label{DIC}
\end{equation}
where $\mathbf{x}$ denotes the respective parameter set of each model and $\Bar{x_i}=\int x_i\mathcal{L}^\mathrm{m}({x_i})\dd x_i$ expresses the mean value of the free parameters. Similarly, $\Bar{\chi}^2=\int\chi^2(\mathbf{x})\mathcal{L}(\mathbf{x})\dd\mathbf{x}$ indicates the mean value of the $\chi^2$ function. Using \eqref{DIC} yields the following results:
\begin{equation}
\mathrm{DIC}_\mathrm{TDiff}=562.05,\hspace{1mm}\mathrm{DIC}_{w\mathrm{CDM}}=561.09,\hspace{1mm}\mathrm{DIC}_{\Lambda\mathrm{CDM}}=562.81;
\label{DICS_obtained}
\end{equation}
where, for completion, we also included the analogous $\Lambda$CDM SNIa+CMB fit to the ones we performed for the other two models (this fit was equivalent to the $w$CDM fit but fixing $w=-1$).

Therefore, we can compute
\begin{equation}
    \Delta\mathrm{DIC}_1=\mathrm{DIC}_\mathrm{TDiff}-\mathrm{DIC}_{w\mathrm{CDM}}=0.96,
    \label{Delta_DIC}
\end{equation}
\begin{equation}
\Delta\mathrm{DIC}_2=\mathrm{DIC}_{\Lambda\mathrm{CDM}}-\mathrm{DIC}_{\mathrm{TDiff}}=0.76;
    \label{Delta_DIC}
\end{equation}
which indicate that there is \textit{weak} evidence in favor of $w$CDM with respect to our TDiff model, since $0\leq\Delta\mathrm{DIC}_1<2$ according to the standards. Similarly, we obtained $0\leq\Delta\mathrm{DIC}_2<2$, and thus in light of this analysis we see that there is \textit{weak} evidence in favor of the TDiff model with respect to $\Lambda$CDM. Consequently, the results obtained in this section indicate that TDiff models provide goodness of fits to CMB and SNIa data statistically similar to those of $w$CDM and future work regarding these models is thus further motivated. Particularly, the development of the perturbation regime and the full observational analysis including BAO, $H(z)$ data, the power spectrum, etc. look to be of special interest.

}

Notice how, although the model is compatible with an approximate cosmological constant behavior in the 2-$\sigma$ region (see Fig.\ref{patricio}), the best fit area lies in the range of $w_2$ in the interval \correct{$(-0.729,-0.677)$}. This indicates that, in light of observational data, the TDiff model would favor a non-cosmological constant behavior in the asymptotic future for the dark energy component, with it being phantom in the matter era and there being phantom crossing, as we studied from the single-field dominance regimes.

\begin{figure}[h]
    \centering
    \includegraphics[scale=0.60]{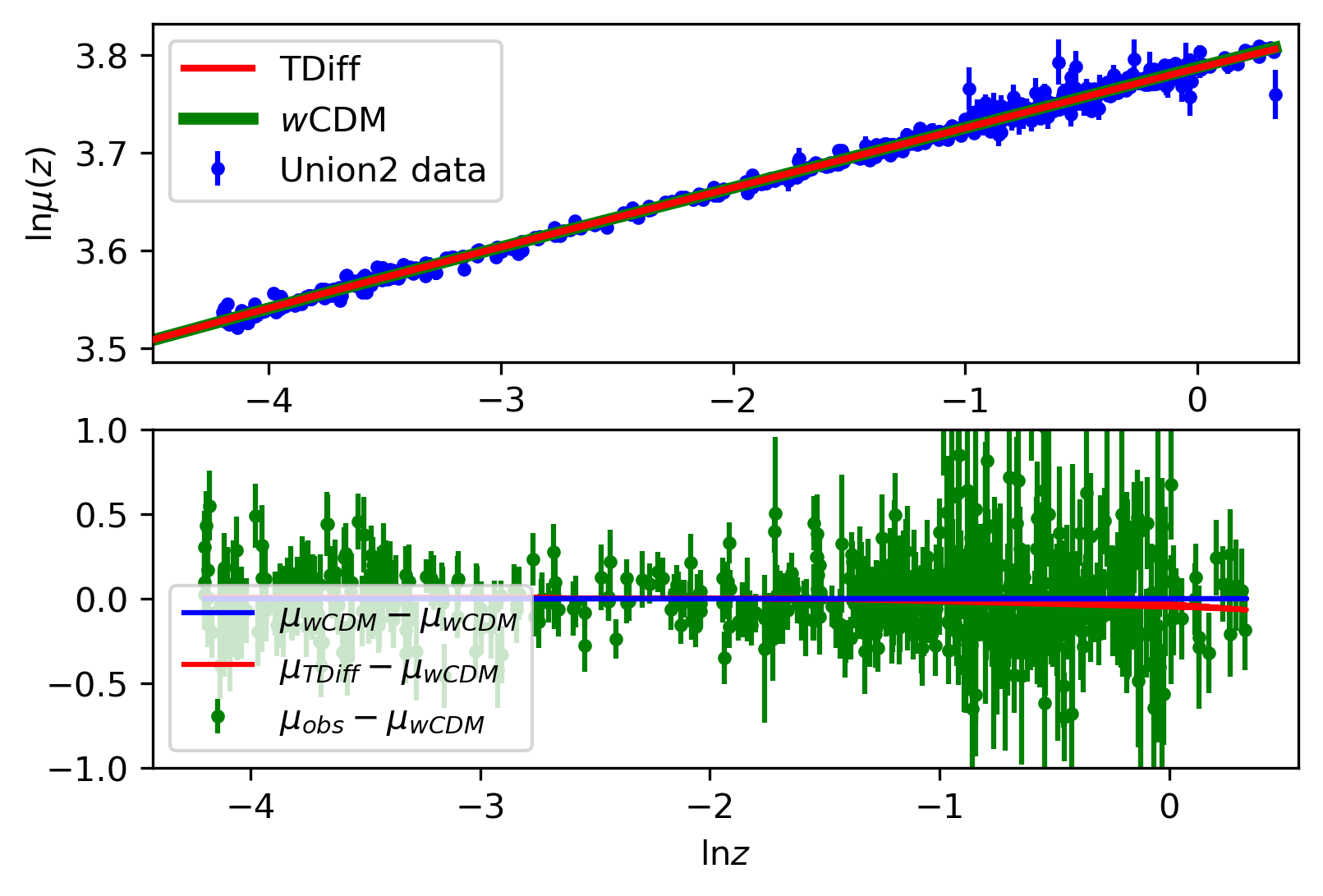}
    \caption{Best fit (SNIa): comparison to $w$CDM and observations. Both models exhibit great accordance with observational data from type Ia supernovae and do not differ much from each other. Minor differences start appearing between the models at higher redshift values.}
    \label{eustaquioguay}
\end{figure}

Fig.\ref{eustaquioguay} summarizes the results of the best fitting SNIa model and its comparison to $w$CDM, displaying favorable agreement with observations, and also to $w$CDM, although there start being minor differences between both models at higher redshift values.

\begin{figure}[h]
    \centering
    \includegraphics[scale=0.60]{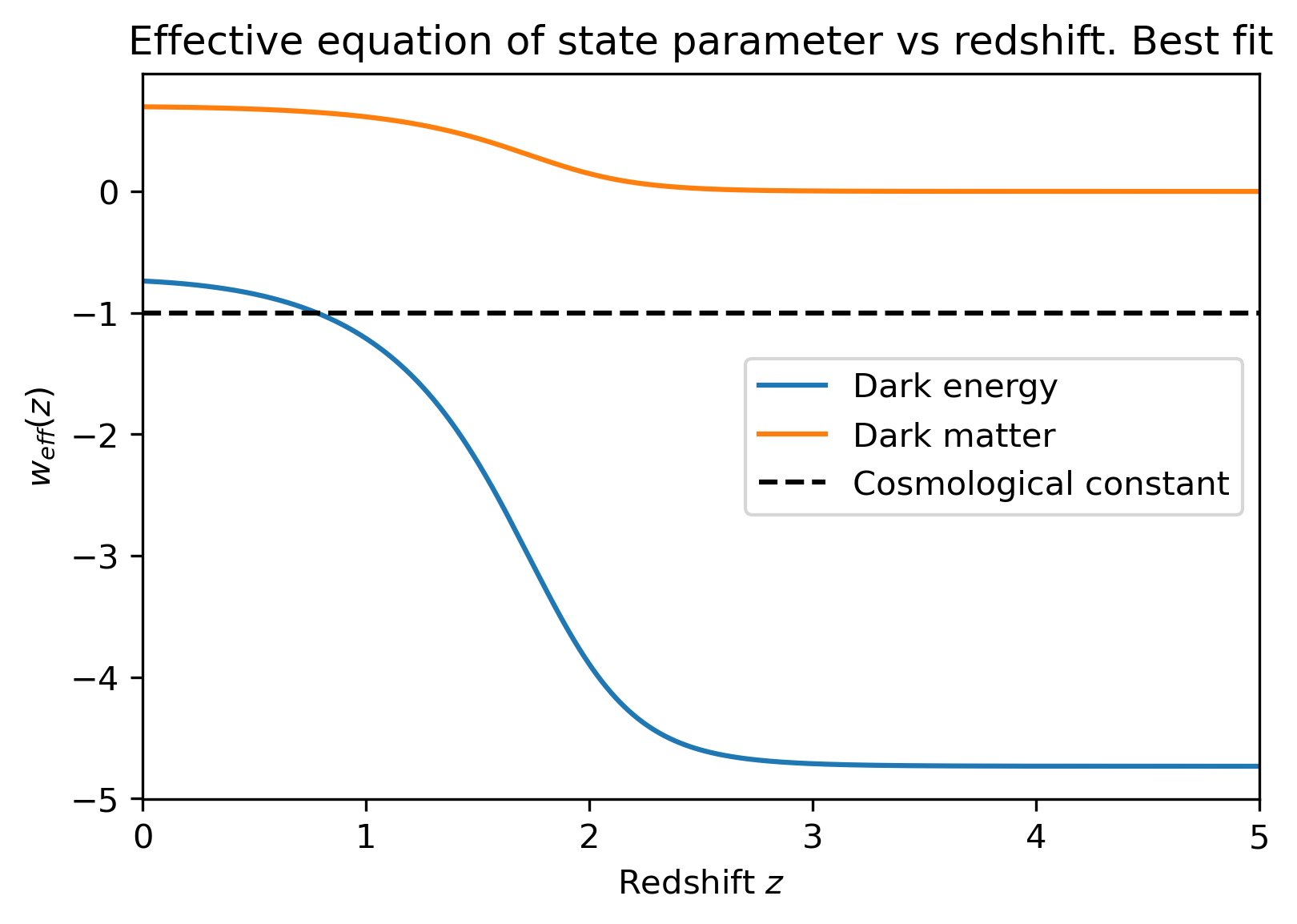}
    \caption{$w_\mathrm{eff}(z)$ for the two-parameter best fit. DE behaves as a phantom component under DM domination and its dynamical decay transitions to depict its quintessence $w_2$ behavior in the future. There is phantom crossing taking place near the current time.}
    \label{Ra}
\end{figure}

Lastly, we include the evolution of the effective equation of state parameter for the DE and DM components for the best fitting TDiff model in Fig.\ref{Ra}.
We see that today DE evolves with an effective equation of state $w_\mathrm{eff,2}(t_0)\simeq -0.75$. As a result, TDiff models favor the presence of a dynamical DE, starting from phantom at early times and slowly transitioning to usual quintessence DE, with an asymptotic quintessence decay dictated by $w_2$). Similarly, DM will exhibit a faster decay than that expected from $w_1=0$ at recent times as a consequence of the symmetry breaking, without the usual $a^{-3}$ decay being altered during the matter era. 

The results obtained throughout this section indicate that this TDiff model should be further explored in the future. Particularly, the Hubble tension problem should be taken into consideration, as other models involving phantom DE have been proven to be favored by observations \cite{DiValentino:2020naf}.

It is worth remarking that this time-evolving DE behavior involving phantom-quintessence transitions was obtained without enforcing any type of interaction potential in the Lagrangian, and without the addition of non-canonical or ghost terms in the matter action.

\section{Conclusions and future work}\label{secV}
In this work, we have considered shift-symmetric theories with two kinetically-driven scalar fields breaking the Diff symmetry down to TDiff and studied their cosmological consequences. We have analyzed the geometrical condition imposed by the conservation of the total EMT, which still holds as a consequence of the Bianchi identities. When working in a flat \correct{FLRW} background, this conservation allows us to obtain a geometrical constraint that leads to a particular shape for the lapse function, which cannot be freely chosen now due to the symmetry breaking. 

This geometrical constraint enforces an exchange of energy between both fields, as their individual EMTs are not conserved. In light of this fact, we have proposed a dark sector model involving two TDiff scalar fields coupled to gravity through power-law functions of the metric determinant, with one field describing a DM fluid and the other DE. We have regarded the different field domination regimes and showed that, although the equation of state parameters of both fluids are constant, both components will exhibit a different dynamical decay as that corresponding to Diff models with the same constant parameters. Particularly, when imposing that DM decays as $a^{-3}$ at early times, we show that the DE component will present phantom behavior during the matter era for it to slowly transition into quintessential behavior in the future, even if its equation of state parameter takes a constant value larger than minus one. It should be emphasized that this interaction of the dark sector is obtained without including interacting potentials. Moreover, non-canonical kinetic terms have not been considered to obtain phantom behaviour. In this framework one naturally obtains an interacting dark sector with a dark energy component that crosses the phantom regime.
In addition, the shift symmetry of the fields allows us to describe a dynamical interacting dark sector avoiding fine-tuning problems depending on the specific choice for the potential. 

We have also studied the evolution of the energy exchange between the fluids and shown how in these models it is always DE which gains energy from DM. On the other hand, we have also considered a particular analytical model to understand the physics involved in this TDiff dark sector framework. For that model, we have analysed the form of the interaction kernel, and investigated the decay of the fluids, parameterized through $w_{\mathrm{eff},i}(z)$.

Beyond the simplest dark sector model, we have studied this interacting dark matter-dark energy framework in deeper detail using numerical techniques. We have considered its parameters: $w_2$ (the equation of state parameter for the DE field, linked to the exponent of the coupling function) and $\Omega_\mathrm{M}^\mathrm{eff}$ (an effective density parameter at high redshift values, which plays a similar role to the $\Omega_\mathrm{M}$ parameter in $\Lambda$CDM). Moreover, we have used the Union2 data for Ia supernovae and fitted our two parameters to these observations to get a first glance regarding the viability of these theories. Our results \correct{show compatibility} with those data and a goodness of the fit similar to that of $w$CDM. \correct{We also studied the statistical performance and checked that, based on our analyses, there is weak evidence that favors $w$CDM with respect to our TDiff model, and there is also weak evidence that favors our TDiff model with respect to $\Lambda$CDM}. It is worth mentioning that these results have been obtained considering two parameters ($w_2$ and $\Omega_\mathrm{M}^\mathrm{eff}$), but further and more complete results should be obtained in future work when extending to the full observational analysis using more data sets and more parameters (namely the Hubble constant and the baryonic abundance, as well as the absolute magnitude). That is, after having introduced this shift-symmetric multi-field TDiff model for the first time in this work and studied it from a more theoretical point of view, these first positive results definitely indicate that the model deserves further analyses.  

Finally, it is worthy to emphasize that the present work has started a new line of research based on studying multi-field TDiff theories, \correct{as we established the theoretical basis for these models and analyzed the interactions arising from the Diff symmetry breaking from a theoretical point of view. We also performed a preliminary observational analysis at the background level that motivates further study of these theories. Consequently,} future projects include to investigate the stability under cosmological perturbations of these theories, in order to study structure formation from the TDiff perturbation formalism perspective; the covariantized approach, studied in detail in reference \cite{Jaramillo-Garrido:2024tdv}, could be of special relevance  for this purpose. Moreover, one could also analyse the models resulting from considering more general coupling functions or non-homogeneous fields, as well as going beyond the shift-symmetric case and/or breaking the Diff symmetry also in the Einstein-Hilbert action. From an observational point of view, future work will also be done regarding a deeper likelihood analysis using additional data sets, such as (Pantheon+SH0Es \cite{Scolnic:2021amr,Brout:2022vxf}, the full CMB likelihood, BAO and $H(z)$ \correct{\cite{RiessAdam,Moresco,Scolnic,zhongguo}}) and an extended parameter space, together with the possible impact on the Hubble tension problem.\\ \\

\textbf{Acknowledgements}\\ \\
The authors would like to thank Darío Jaramillo Garrido and Alfredo Delgado Miravet for useful comments
and discussions regarding TDiff theories. DTB also acknowledges financial help from the Ayudas
de Máster IPARCOS-UCM/2023.
This work has been supported by the MICIN (Spain) Project No.
PID2022-138263NB-I00 funded by MICIU/AEI/10.13039/501100011033 and by
ERDF/EU.


\bibliography{bibliografia} 
\bibliographystyle{elsarticle-num}

\end{document}